\documentclass[11pt,a4paper]{article}
\pdfoutput=1 
\usepackage{jheppub}	

\usepackage{tikz}
\usetikzlibrary{patterns}
\usetikzlibrary{positioning,arrows}
\usetikzlibrary{decorations.pathmorphing}
\usetikzlibrary{decorations.markings}
\usetikzlibrary{snakes}
\usetikzlibrary{matrix}

\usepackage{lmodern}

\usepackage{amsmath}
\usepackage{bm}
\usepackage[utf8]{inputenc}
\usepackage[english]{babel}
\usepackage{makeidx}
\usepackage{tikz}
\tikzset{every picture/.style={line width=0.75pt}} 
\usepackage{amsfonts}
\usepackage{enumerate}
\usepackage{mathrsfs}
\usepackage{tensor}
\usepackage[autostyle]{csquotes}
\usepackage{subfig}

\def\be{\begin{equation}}
	\def\ee{\end{equation}}
\def\bea{\begin{eqnarray}}
	\def\eea{\end{eqnarray}}
\newcommand{\nn}{\nonumber}

\newcommand{\ft}[2]{{\textstyle\frac{#1}{#2}}}

\def\apjl{\ref@jnl{ApJ}}

\newcommand{\RNum}[1]{\uppercase\expandafter{\romannumeral #1\relax}}

\usepackage{amsmath,amssymb}
\usepackage{latexsym}
\usepackage{graphicx}
\usepackage{slashed}

\usepackage[vcentermath,enableskew]{youngtab}
\usepackage{epsfig}

\def\be{\begin{equation}}
	\def\ee{\end{equation}}
\def\bea{\begin{eqnarray}}
	\def\eea{\end{eqnarray}}

\title{A note on rank $\frac{3}{2}$ Liouville irregular block }
\author[]{Rubik Poghossian and}
\author[]{Hasmik Poghosyan }
\emailAdd{poghos@yerphi.am}
\emailAdd{h.poghosyan@yerphi.am}

\affiliation[]{Yerevan Physics Institute \\
	Alikhanian Br. 2, 0036 Yerevan, Armenia}

\abstract{This paper focuses on a conformal block  with rank $\frac{3}{2}$ irregular singularity 
	which corresponds to the prepotential of the  ${\cal H}_1$ Argyres-Douglas theory in $\Omega$ background.
	We derive this irregular conformal block  using generalized holomorphic anomaly recursion relation. This results is an expression 
	which is 	a power series in $\Omega$-background parameters  $\epsilon_{1,2}$ and exact in coupling.
	We have verified that in small coupling regime   our result is consistent with previously known expressions.
	
	Furthermore we derive the Deformed Seiberg-Witten  curve which provides an alternative tool to explore above 
	mentioned theory in Nekrasov-Shatashvili limit of $\Omega$-background.
	We checked that the results are in complete agreement with the holomorphic anomaly approach.
}
 \clearpage
\begin{document}
	\tikzset{
		line/.style={thick, decorate, draw=black,}
	}
	
	\maketitle
	

\section{Introduction}
Equivariant localization is a powerful technique to investigate supersymmetric theories.  In particular 
its application to   Seiberg-Witten (SW) theory  \cite{Seiberg:1994rs,Seiberg:1994aj} allows one to reduce seemingly intractable instanton 
moduli integrals to essentially smaller submanifold of instantons invariant with respect to an Abelian  
subgroup of global gauge singled out by the adjoint scalar VEV \cite{Bellisai:2000bc,Flume:2001kb,Flume:2001nc,Dorey:2002ik}. Embedding the theory in a specific 
graviphoton (so called $\Omega $-) background leads to further drastic simplifications. From the localization 
point of view this amounts to incorporating (Euclidean) space-time rotations on two orthogonal planes parameterized 
by the angles $\epsilon_1$, $\epsilon_2$ with global gauge transformations. As a result, the integrals mentioned above localize in 
each instanton sector to a finite number of fixed points, which are labeled by Young diagrams.
One recovers the initial SW theory by  
simply sending the $\Omega$-background parameters $\epsilon_{1,2}$ to zero.
For explicit computations, various localization techniques have been developed
\cite{Nekrasov:2002qd,Flume:2002az,Bruzzo:2002xf,Nekrasov:2003rj,Flume:2004rp}.

Over time, it has been realized that the partition function of the gauge theory in 
$\Omega$-background can be mapped into correlation functions of two-dimensional conformal field theory (CFT) 
\cite{Gaiotto:2009we,Alday:2009aq,Poghossian:2009mk}. This remarkable correspondence is referred as AGT 
(Alday, Gaiotto,  Tachikawa) duality.

On the other hand it was demonstrated by Argyres and Douglas  \cite{Argyres:1995jj} three decades ago, that 
there are specific points in the moduli space of certain ${\cal N}=2$ supersymmetric gauge theories, 
commonly referred as Argyres-Douglas points, where both electrically and magnetically charged particles 
become massless simultaneously.

It appears that the partition functions of Argyres-Douglas theories in presence of  
$\Omega$-background, from 2d CFT point of view corresponds  
to conformal blocks of so called irregular states.
These new class of states  (called rank $r$ irregular states) were
introduced in \cite{Gaiotto:2009ma,Gaiotto:2012sf},
where a systematic method was developed 
for constructing integer rank $r$ irregular states in 2d Liouville  CFT 
by colliding insertion points of $r+1$ primary fields. Detailed investigations of relations between 
irregular blocks  and AD theories were further carried out in
\cite{Nishinaka:2012kn,Kanno:2013vi,Bonelli:2016qwg,Nishinaka:2019nuy,Kimura:2020krd,
	Bonelli:2021uvf,Bonelli:2022ten,Kimura:2022yua,Consoli:2022eey,Fucito:2023plp}.

In \cite{Poghosyan:2023zvy} the simplest AD theory ${\cal H}_0$ related 
to rank $\frac{5}{2}$ conformal block, was included coherently into 
above scheme and in-depth analysis was carried out for the corresponding irregular conformal block. 
This allowed  \cite{Hamachika:2024efr} to give    
a definition for arbitrary half integer rank irregular states.

The focus of this paper is the rank $\frac{3}{2}$ irregular conformal block 
\be
\label{ZH1_def}
{\cal G} =\langle  \Delta_{\beta_0}  |I^{(3/2)}(c_1,\Lambda_3;\beta_0,c_0)\rangle\,,
\ee
which is related to the 
${\cal H}_1$ Argyres-Douglas theory in $\Omega$ background. Here $|\Delta_{\beta_0} \rangle$ is
Liouville  primary state with dimension
\bea
\Delta_{\beta_0}=\beta_0(Q-\beta_0)\,,
\eea
where
\bea
Q=b+\frac{1}{b}
\eea
parametrises the Virasoro central  charge 
\bea
c=1+6Q^2\,.
\eea
Notice that we have chosen both  primary  and irregular state parameters to take the same value  $\beta_0$,
 otherwise from standard CFT arguments the irregular block would vanish.
Using the holomorphic anomaly recursion relation \cite{Huang:2006si,Huang:2009md,Huang:2011qx,Huang:2013eja} 
we have found this irregular conformal block up to order $8$ in $\epsilon_{1,2}$. It is important to 
emphasize that the less standard form of its discriminant made the application of the recursion relation 
more complicated.
Our result (\ref{F2}) after expanding it in $\Lambda_3$ perfectly match with those obtained through 
irregular state computation carried out in \cite{Hamachika:2024efr} or (\ref{v_cft}).

In \cite{Poghossian:2010pn,Fucito:2011pn,Nekrasov:2013xda}, the concept of a so-called Deformed Seiberg-Witten (DSW) curve was introduced 
(see also \cite{Poghosyan:2020zzg}). This method provides an efficient method to investigate gauge theories in 
particular case of $\Omega$-background when one of the deformation parameters, say $\epsilon_1=0$ (Nekrasov-Shatashvili 
limit \cite{Nekrasov:2009rc}). We have derived DSW curve equation for our present case of interest the ${\cal H}_1$ theory and then used it to 
derive the deformed prepotential and corresponding heavy conformal block.

This paper is structured as follows:
In Section \ref{irr_states}, we provide a review of the $r = \frac{3}{2}$ irregular state and its corresponding conformal block.
Section \ref{hol_recursion} begins with an overview of the holomorphic anomaly recursion relation. 
In Subsection \ref{H1_SW_exact}, we outline the framework for applying this recursion to our specific case.
Our results are presented in Subsection \ref{secHolAnAppl}.
In Section \ref{HeavyCFTblock}, we derive the analog of the Belavin–Polyakov–Zamolodchikov (BPZ) differential equation for rank $3/2$,
and subsequently explore its heavy limit. From this differential equation, we determine the DSW curve, which enables us to derive the heavy conformal block.  
The appendices provide detailed computations and results that were too extensive to be included in the main text.
\section{Rank $\frac{3}{2}$ irregular state and its corresponding conformal blocks}
\label{irr_states}
We use the following notation for  rank $3/2$ irregular state 
\bea
\label{Irr_st_par}
|I^{(3/2)}( c_1,\Lambda_3;\beta_0,c_0)\rangle
\equiv
|I^{(3/2)}\rangle\,.
\eea
This states are defined as 
\bea
\label{def_ir32}
L_k |I^{(3/2)}\rangle  &=& 
{\cal L}^{(3/2)}_k|I^{(3/2)}\rangle\,; ~~ \qquad \qquad \quad k=0,\ldots,3\,, \\
L_k |I^{(3/2)}\rangle  &=&0 \,;  ~~ \qquad \qquad\qquad\qquad \qquad k>3\,,
\eea
with
\bea
\nn
{\cal L}^{(3/2)}_0 =3 \Lambda_3 \frac{\partial}{\partial \Lambda_3}+c_1 \frac{\partial}{\partial c_1}\,,
\qquad
{\cal L}^{(3/2)}_1 =-\frac{\Lambda_3}{2 c_1}  \frac{\partial}{\partial c_1} \,, \\
{\cal L}^{(3/2)}_2 =- c_1^2\,,
\qquad
{\cal L}^{(3/2)}_3  =\Lambda_3\,.
\label{ln_32}
\eea
This operators are  constructed  to be compatible with the Virasoro algebra. 
It is natural to call them rank $3/2$ states since they demonstrate characteristics that are intermediate between rank  
 $1$ and rank $2$ (the definitions of integer rank states can be found  in appendix \ref{IntegerRankIrregularStates}).

We will be interested in the  irregular conformal block (\ref{ZH1_def})
which is related  to the partition function of  ${\cal H}_1$ Argyres-Douglas theory in $\Omega$ background.
We can see this by
following \cite{Alday:2009aq} where one computes the (normalized) 
expectation value of the Liouville stress-energy tensor
\be
\label{GaiDif}
\phi_2(z)   =- { \langle \Delta_{\beta_0} | T(z) | I^{(3/2)} \rangle  \over \langle \Delta_{\beta_0}   | I^{(3/2)}\rangle }
\ee
and identifies the $1$-form $\sqrt{\phi_2(z)} dz$ with the SW differential (see bellow).

From (\ref{GaiDif}) using (\ref{def_ir32})  one gets
\bea
\label{GswCurve}
\phi_2(z)  =
-{\Delta_{\beta_0} \over z^2}  {+}{ 2 v \over z^3}
{+}{c_1^2\over z^4}
{-}{\Lambda_3 \over z^5} \,,
\eea
with
\be
\label{G_Matone_CFT}
v=\frac{\Lambda_3}{4c_1}\,\partial_{c_1}\log{\cal Z}_{{\cal H}_1} \,.
\ee
It is the form of (\ref{GswCurve}) which suggests that  we are dealing with the   
${\cal H}_1$ Argyres-Douglas theory in $\Omega$ 
background\footnote{Indeed by re-scaling 
$ z \to \Lambda_3 ^ {\frac{1}{3}} z $, we can set the coefficient at the 
leading singularity $z^{-5}$ of  $\phi_2 d z^2$ to $1$.  
Then, equating the dimension $\left[ \phi_2 d z^2 \right]  =2 $ 
we observe that the new coefficients identified as the relevant coupling 
and the respective field have dimensions $\left[\frac{2}{3}\right]$ and 
$\left[\frac{4}{3}\right]$.}. 
	The partition function ${\cal Z}_{{\cal H}_1}$ was derived as a series in 
$\Lambda_3$ in \cite{Hamachika:2024efr}. We present the details of computation 
in appendix \ref{ZH1LambdaExpansion}.
Inserting (\ref{Zh1inst}) in (\ref{G_Matone_CFT}) we get
\bea
\label{v_cft}
v&=&c_1 \left(c_0-Q\right)+\frac{3 \left(c_0-Q\right){}^2+\Delta_{\beta_0}}{4 c_1^2}\Lambda _3
\\ \nn
&-&\frac{\left(c_0-Q\right) \left(34 \left(c_0-Q\right){}^2+18 \Delta_{\beta_0}+5 Q^2-1\right)}{32 c_1^5}\Lambda _3^2 
+\ldots\,.
\eea
To make contact with standard  Seiberg-Witten curve analysis,
we introduce gauge theory parameters as follows: 
\bea
\label{map_CFT_H1} 
&& Q=\frac{s}{\sqrt{p}}\,,
\quad
c_0=\frac{\hat{a}+s}{\sqrt{p}}\,,
\quad
v ={\hat{v} \over p} \,,
\\ \nn
&& \phi_2=\frac{\hat\phi_2}{p}\,,
\quad
\Lambda_3=\frac{\hat\Lambda_3}{p}\,,
\quad
k=\frac{{\hat k}}{\sqrt{p}}\,,
\quad
c_1=\frac{\hat{c}_1}{\sqrt{p}}\,,
\eea
where 
\bea
\label{s_p_def}
s=\epsilon_1+\epsilon_2 \,,
\qquad
p=\epsilon_1\epsilon_2\,
\eea
and for convenience sake instead $\beta_0$ we will use the momentum parameter $k$
\bea
\beta_0=\frac{Q}{2}-k \,.
\eea
Thus we have
\bea
\label{phi2hat}
\hat\phi_2(z) &=&{\hat{	k}^2 \over z^2}  {+}{ 2 \hat{v} \over z^3}
{+}{\hat{c}_1^2\over z^4}
{-}{\hat{\Lambda}_3 \over z^5} \,.
\eea
The Seiberg-Witten  differential is given by
\be 
\label{SW_dif}
\lambda_{SW}=\sqrt{{\hat \phi}_2(z)}\, \frac{dz}{2 \pi i}
\ee 
and the $A$-cycle is defined as
\bea
\label{Acycle}
\hat{a}=\oint_{C_A} \lambda_{SW}\,.
\eea
  Notice, that $A$-cycle shrinks to 
the point $z=0$ in ${\hat \Lambda_3}\to 0$ limit, so that in this case to evaluate the contour 
integral (\ref{Acycle}) one can simply expand
$\sqrt{\hat\phi_2}$ in powers of $ \hat{\Lambda}_3$ and then take the residues at $z=0$. 
The resulting expression is given by
\bea
\label{aSWflat}
{\hat a }=
\frac{\hat v}{\hat c_1}-\frac{3 \hat v^2- {\hat c}_1^2 \hat{k}^2}{4 \hat c_1^5}\hat\Lambda _3
+\frac{5  \hat v \left(7 \hat v^2-3 \hat c_1^2\hat{k}^2 \right)}{16 \hat c_1^9}\hat \Lambda _3^2
+O(\hat{\Lambda}_3^3)\,.
\eea
Inverting for $\hat{v}$ one finds
\bea
\label{Flatv}
\hat{v}={\hat a} {\hat c}_1
+\frac{ 3 {\hat a}^2-\hat{k}^2}{4 {\hat c}_1^2}{\hat \Lambda} _3
 -\frac{{\hat a}  \left(17 {\hat a}^2-9 \hat{k}^2\right)}{16 {\hat c}_1^5}{\hat \Lambda} _3^2
 +O(\hat{\Lambda}_3^3)\,.
\eea
Except $\epsilon_{1,2}$-corrections this nicely matches (\ref{v_cft}) by
taking into account (\ref{map_CFT_H1}). 

\section{ The holomorphic anomaly recursion }
\label{hol_recursion}
In this section, we will derive a formula for the prepotential of ${\cal H}_1$
Argyres-Douglas theory, which is exact with respect to the coupling constant but
is a perturbative expansion in $\Omega$-background parameters. Note that the CFT
 approach discussed in previous section offers a complementary 
perspective: it produces a power series in coupling with coefficients that are exact in 
$\epsilon_1$ and $\epsilon_2$.

\subsection{The SW  prepotential in flat background}
Let us  consider any SW theory governed 
by an elliptic curve. 
Assume that this elliptic curve is expressed in its Weierstrass canonical form
\be
\label{WeiElCu}
y^2 =4 z^3 - g_2 z- g_3\,,
\ee
where $g_2$ and $g_3$ are polynomials in global modulus parameter $u$.
The periods of the Weierstrass elliptic curve are defined as follows:
\be
\omega_i =\oint_{\gamma_i} \frac{dz}{\pi y} \,,
\ee 
where  $\gamma_1$ and $\gamma_2$ are appropriately chosen
$A$ and $B$ cycles of the torus. 
As is customary, the infrared coupling  $\tau_{IR}$ is defined in terms of the torus parameters as follows:
\bea
\tau_{IR}= {\omega_2\over \omega_1}\,.
\eea
For convenience sake, we introduce the nome, defined by
\bea
q=e^{\pi {\rm i}\tau_{IR}}\,.
\eea
Due to standard formulae  of elliptic geometry we have
\bea
g_2=\frac{4}{3\omega_1^4}E_4(q)\,,\qquad 
g_3=\frac{8}{27\omega_1^6}E_6(q)\,,
\eea 
where the Eisenstein series are defined by ($k=2,4,6,\cdots $)
\bea
E_k(q)=
1+{2\over \zeta (1-k)}\sum _{n=1}^{\infty} \frac{n^{k-1} q^{2 n}}{1-q^{2 n}}\,.
\eea
Notice that the  following relationships hold
\bea
\label{u_q_relation}
{27 g_3^2\over g_2^3} = { E_6(q)^2\over E_4(q)^3} \, , 
\qquad
\omega_1(q,u)^2= \frac{2g_2E_6(q)} {9 g_3E_4(q)} \,.
\eea
From the first equation, we can derive:
\be
\label{qdqu}
D_\tau u\equiv q \partial_q u  = {2\left(E_4^3-E_6^2\right) 
	\over E_4 E_6 \left( {3g_2'(u) \over g_2(u)} -
	{2g_3'(u)  \over g_3(u) } \right)}\,,
\ee
here we have  applied Ramanujan's differentiation rules:
\be
\label{qdqE46} 
\quad D_\tau E_4 = \ft{2}{3} (E_2 \, E_4-E_6) \, , \quad
D_\tau E_6=  E_2 \, E_6-E_4^2\,.
\ee
For our purposes, it is also important to recall the differentiation rule for the degree 2 quasi-modular form:
\be
\label{qdqE2}  
D_\tau E_2 = \ft{1}{6} (E_2^2-E_4)\,.
\ee
The "flat" coordinate $a$, $a_D$ and the SW prepotential ${\cal F}_0(a)$ are 
introduced through standard relations 
\bea
\label{defF0}
-\frac{1}{2}{\cal F}_0''(a)&=&\log q \, , \\
\label{def_aAndaD}
\omega_1(q,u)&=&\partial_u a \,\,,
\qquad
\omega_2(q,u)=\partial_u a_D\,.
\eea 
Following \cite{Huang:2006si,Huang:2009md,Huang:2011qx,Huang:2013eja} we introduce the quantities
\be
S={2\over 9\omega_1(q,u)^2 }=\frac{g_3(u)E_4(q)}{g_2(u)E_6(q)}  \, ,\qquad  X = S  E_2(q)\,.
\ee
Their total $u$-derivatives can be computed using the equations (\ref{qdqu}), 
(\ref{qdqE46}), (\ref{qdqE2}). Here are the results:
\begin{small}
	\bea
	p_1 &=& {d\over du} \ln S =\frac{\Delta '(u)}{6 \Delta (u)}+\frac{9 X w (u)}{2 \Delta (u)}\,, \\
	\label{p2}
	p_2 &=&{dX\over du}  =\frac{g_2(u) w (u)}{12 \Delta (u)}+\frac{9 X^2 w (u)}{4 \Delta (u)}+\frac{X \Delta '(u)}{6 \Delta (u)}\,,
	\eea
\end{small}
where
\bea
w(u)=2 g_2(u) \frac{\partial g_3(u)}{\partial u}-3 g_3(u) \frac{\partial g_2(u)}{\partial u}\,.
\eea
It is easy to check that the derivatives of $a$ with respect to $u$, in terms of quantities 
introduced above,  are given by
\bea
\label{uarel}
\left(\frac{du}{da}\right)^2 &=& {1 \over  \omega_1^{2} } 
=\frac{9S}{2}\, , \qquad \frac{d^2u}{da^2} =\frac{1}{2}\frac{d}{d u}\, \omega_1^{-2}
=\frac{9S\,p_1}{4} \,.
\eea
\subsection{Holomorphic anomaly equation}
\label{holAnRecRel}
In terms of $\Omega$-background parameters 
\[
s=\epsilon_1+\epsilon_2 \,, \qquad p=\epsilon_1 \epsilon_2\,
\]
the full prepotential can be represented as a double power series
\bea
{\cal F}=p \log Z=\sum_{n=0,m=0}^\infty s^{2n}
p^{m}F^{(n,m)} =\sum_{g=0}^{\infty}
p^{g}{\cal F}_g\,,
\eea
where
\be
{\cal F}_g=\sum_{n+m=g}\left({s^{2}\over p}\right)^nF^{(n,m)}\,.
\ee
The term ${\cal F}_0$ which does not  depend on $\epsilon_{1,2}$ is just the SW prepotential. 

Higher order terms ${\cal F}_g$ with $g>1$ can be computed recursively using holomorphic anomaly equation
\be\label{HAE}
\partial_{E_2} {\cal F}_{g}  =\ft{1}{24} \left(    \partial_a^2 {\cal F}_{g-1} +\sum_{g'=1}^{g-1} \partial_a {\cal F}_{g'} \partial_a {\cal F}_{g-g'}  \right)\,,
\ee
starting from $g=1$ expression
\be
\label{f1h1}
{\cal F}_1 (u,b,q)={1 \over 4}  \log {1\over\omega_1(q,u)^2} 
+\ft{ s^2-2p }{24p}\log \Delta (u) \,,
\ee
where
\be
\label{disc}
\Delta(u)=g_2^3
-27g_3^2
\ee 
is the modular discriminant.
Using (\ref{uarel}) one can rewrite (\ref{HAE}) as
\bea
\label{HAE2}
\partial_{X} {\cal F}_{g}
&=&  \frac{3}{16} \left( D_u^2  {\cal F}_{g-1}+ {p_1\over 2} D_u  {\cal F}_{g-1}
+  \sum_{g'=1}^{g-1}  D_u  {\cal F}_{g'}D_u  {\cal F}_{g-g'} \right)\,.
\eea
In this setting one should consider ${\cal F}_{g}$ as a function of two  variables 
	$u$ and $X$. The total derivative with respect to $u$, $D_u$, due to  (\ref{p2}) is
\bea 
D_u=\frac{\partial}{\partial u}+p_2\frac{\partial}{\partial_X}\,.
\eea 	
A careful analysis carried out in \cite{Huang:2011qx} shows that
${\cal F}_{g}$ is a polynomial in $X$ of maximal degree $3(g-1)$ with rational in $u$ 
coefficients. More precisely the denominators of this coefficients are equal 
to $\Delta(u)^{2g-2}$ and numerators are polynomials in $u$ of degree smaller than 
$2d_{\Delta}(g-1)$. Thus
\bea
\label{HA}
{\cal F}_{g}|_{X=0}=\Delta(u)^{2-2g}\sum_{i=0}^{2d_{\Delta}(g-1)-1}s_i u^i\,,
\eea
where $d_{\Delta}$ is the degree of discriminant in $u$, $s_i$ are some unknown coefficients.
Evidently, the equation (\ref{HAE2}) alone can not fix $X$ independent terms. This ambiguity 
can be removed imposing so called gap condition. Namely, for $g>1$ near each zero $u_*$ 
of the discriminant, the gap condition reads
\bea\label{gap_condition}
{\cal F}_g  & \underset{ u\to  u^*} {\approx}  &  {\frac{(2g-3)!}{a^{2g-2}}  \sum_{k=0}^{g} \hat B_{2k} \hat B_{2g-2k}\left(\frac{\epsilon_1 }{\epsilon_2}\right)^{g-2k}}  + O(a^0)\,,
\eea
where
\be
\hat{B}_m=\left( 2^{1-m}-1\right) {B_m\over m!}\,,
\ee
with $B_m$ the Bernoulli numbers and $a$ is the local flat coordinate, vanishing 
at $u=u^*$. 
Notice the absence of lower order poles 
$a^{-n}$ with $n<2g-2$ in (\ref{gap_condition}), hence the term ``gap condition". 
\subsection{${\cal H}_1$ in flat background}
\label{H1_SW_exact}
Our starting point is the Seiberg-Witten differential (\ref{SW_dif}).
For our case  $\hat{\phi}_2$ is given by (\ref{phi2hat}). One can bring this  curve into canonical form 
by performing the following change of variable
\be 
z=\frac{3 \hat{\Lambda} _3}{\hat{c}_1^2+3 x}\,.
\ee 
Our  SW differential becomes
\be 
\label{SW_dif_canonical}
\hat{\lambda}_{SW}=\frac{3 \sqrt{-4 x^3+g_2 x +g_3}}{2 \hat{\Lambda} _3 \left(\hat{c}_1^2+3 x\right)}\, \frac {dx}{2 \pi i}\,,
\ee
with Weierstrass parameters
\be 
\label{g2_g3_H0}
g_2=\frac{4}{3}   \left(6 \hat{\Lambda} _3 {\hat v}+ \hat{c}_1^4\right)\,, \quad 
g_3=\frac{4}{27} \left(18 \hat{c}_1^2 \hat{\Lambda} _3 {\hat v}+2 \hat{c}_1^6+27 \hat{k}^2  \hat{\Lambda} _3^2\right)\,.
\ee 
Notice that in previous sections we have used  $u$ and $a$ instead of $\hat{v}$ and ${\hat a}$.
For the holomorphic differential we get
\be 
\partial_{{\hat v}}\hat{\lambda}_{SW}=\frac{2}{\sqrt{-4x^3+g_2x+g_3}}\,\frac{dx}{2 \pi i}\,.
\ee
The periods of this holomorphic differential can be expressed in terms of 
the hypergeometric function
\bea 
\label{dua}
\partial_{{\hat v}}{\hat a}&=&
\left(\frac{3 g_2}{4}\right)^{-\frac{1}{4}}\,
_2F_1\left(\frac{1}{6},\frac{5}{6};1;\frac{1}{2}-\frac{1}{2}\sqrt{\frac{27 g_3^2}{g_2^3}}\right)\,,\\
\label{duad}
\partial_{{\hat v}}{\hat a}_D&=&i \left(\frac{3 g_2}{4}\right)^{-\frac{1}{4}} \,
_2F_1\left(\frac{1}{6},\frac{5}{6};1;\frac{1}{2}+\frac{1}{2}\sqrt{\frac{27 g_3^2}{g_2^3}}\right)\,.
\eea 
Expanding (\ref{dua}) for small  $\Lambda_3$ and integrating over ${\hat v}$, up to ${\hat v}$-independent terms one easily recovers   
(\ref{aSWflat}) . With some efforts it is possible to show that 
the ${\hat v}$ independent part also admits a closed expression:
\bea
{\hat a}|_{{\hat v}=0}=\frac{{\hat k}^2 {\hat \Lambda} _3 \, }{4 {\hat c}_1^3}
\, _3F_2\left(\frac{1}{2},\frac{5}{6},\frac{7}{6};\frac{3}{2},2;-\frac{27 {\hat k}^2 {\hat \Lambda} _3^2}{4 {\hat c}_1^6}\right)\,.
\eea 
Notice that due to (\ref{def_aAndaD})
\bea
\label{q} 
q=\exp \left(\pi i \frac{\partial_{{\hat v}} {\hat a}_D}{\partial_{{\hat v}} {\hat a}}\right)\,.
\eea
Now it is straightforward to expand  $q$ in ${\hat \Lambda}_3$ the result is
\bea
\label{qSer}
q^2=\frac{ {\hat a}^2-{\hat k}^2}
{64 {\hat c}_1^6}{\hat \Lambda} _3^2-\frac{ {\hat a} \left(17 
	{\hat a}^2-19  {\hat k}^2\right)}
{128 {\hat c}_1^9}{\hat \Lambda} _3^3+O({\hat \Lambda}_3^4)\,.
\eea 
Using (\ref{defF0}) one can  derive the SW prepotential  
in flat background up to  terms that are independent or linear in ${\hat a}$. 
The contribution of these terms 
can be  partially recovered  (besides possible terms depending solely on ${\hat c}_1$) 
by   using  (\ref{G_Matone_CFT}), (\ref{Flatv}). Below is the result up to second order
    \bea
	\label{F0ser}
	{\cal F}_0&=&\left(3 {\hat a}^2-\hat{k}^2\right) \log {\hat c}_1
	-\frac{1}{2} \left(\hat{k}^2+{\hat a}^2\right) \log \left({\hat a}^2-\hat{k}^2\right)
	\\ \nn
	&+&{\hat a}^2 \left(\frac{3}{2}-\log \frac{{\hat \Lambda}_3}{8}\right)
	+\frac{4 {\hat a} {\hat c}_1^3}{3 {\hat \Lambda}_3}
	-2 {\hat a} \hat{k} \log \frac{\hat{k}+{\hat a}}{\hat{k}-{\hat a}}
	\\ \nn
	&+&\frac{{\hat a}  \left(17 {\hat a}^2-9 {\hat k}^2\right)}{12 {\hat c}_1^3}{\hat \Lambda}_3
	-\frac{ \left(375 {\hat a}^4 -258 {\hat a}^2 {\hat k} ^2+11 {\hat k} ^4\right)}{192 {\hat c}_1^6}{\hat \Lambda}_3^2
	+O\left({\hat \Lambda}_3^3\right)\,.
	\eea
\subsection{Applying holomorphic anomaly recursion to the ${\cal H}_1$ theory}
\label{secHolAnAppl}
Here we apply the method described in previous subsection \ref{holAnRecRel} for the case of 
our main interest ${\cal H}_1$ theory.  
We will  derive $\Lambda_3$-exact formulae for the first few ${\cal F}_g$-terms using  the holomorphic recursive algorithm. 
We  know that
\begin{itemize}
	\item 
    ${\cal F}_0$ can be derived using (\ref{defF0}), (\ref{dua}), (\ref{duad}), (\ref{q}).
	\item ${\cal F}_1$ is given by (\ref{f1h1}).
	\item ${\cal F}_g$ with $g=2,3,\ldots$ can be derived using (\ref{HA}) and (\ref{gap_condition}).
\end{itemize} 
From (\ref{g2_g3_H0}), (\ref{disc}) for the discriminant  we get
\bea
\label{disc_exp}
\Delta(\hat{v})=-16 {\hat \Lambda} _3^2 \left(36 {\hat c}_1^2 {\hat k}^2 {\hat \Lambda} _3 \hat{v}+4 {\hat c}_1^6 {\hat k}^2-4 {\hat c}_1^4 \hat{v}^2+27 {\hat k}^4 {\hat \Lambda} _3^2-32 {\hat \Lambda} _3 \hat{v}^3\right)\,.
\eea
Obviously (\ref{HAE2}) determines all $X$ dependent terms of ${\cal F}_2$.
The more intricate task is identification of coefficients $s_i$ entering in (\ref{HA}).
To accomplish this, we have utilized the gap condition (\ref{gap_condition}) without relying on an explicit knowledge of the roots of discriminant (\ref{disc_exp}) \footnote{If the roots of $\Delta(u)=0$ can be found without using radicals, imposing gap condition is quite straightforward. In our case the discriminant  $\Delta(u)$ is a generic third order polynomial in $u$ which makes our task  more difficult. We have used the algebraic approach suggested in \cite{Huang:2011qx}. }.
Here is our result

\begin{tiny}
	\bea
	\nn
	 {\cal F}_2=\frac{ 9}{2^{10} \Delta^2}
	 \bigg[
	 45 X^3 w^2
	 +
	 X^2 \left(\left(2 Q^2-11\right)w \Delta '+12 \Delta w '\right)+
	X \left(2 g_2 w^2+\frac{1}{27} \left(Q^2-1\right) \left(\left(Q^2-23\right) \Delta '^2+24 \Delta  \Delta ''\right)\right)+
	\\ \label{F2}
\frac{64 \Lambda _3^2}{135}\bigg(
\Delta  \left(9 g_3 \left(53 Q^4-142 Q^2+71\right)-2 c_1^2 g_2 \left(67 Q^4-178 Q^2+89\right)\right)
+g_3 \left(237 Q^4-618 Q^2+299\right) \left(2 c_1^2 g_2-9 g_3\right){}^2
\bigg)
	 \bigg].
	\eea 
\end{tiny}
Upon expanding these expression in small $\hat{\Lambda}_3$, we have verified its  consistency  with (\ref{v_cft}) and (\ref{Aheavy}).

It is interesting to note, that our result for $X$ independent term of ${\cal F}_2$ completely agrees 
with conjectural general expression suggested in \cite{Bonelli:2025owb}

\begin{scriptsize}
	\bea
	\label{BST}
&	{\cal F}^{conj}_2|_{X=0}=\frac{2^{-8}}{5 \Delta^2}
	\left(
	(9-18 Q^2+7 Q^4)\Delta w^{'}g_2-
	\frac{71-142 Q^2+53 Q^4}{12}\Delta^{'} w g_2+
	\frac{3}{2}(43-96 Q^2+39 Q^4)w^2 g_3
	\right)\,.\qquad
	\eea
\end{scriptsize}
Now we can proceed to calculate the next term 
	\bea
	\label{F3_fi_rel}
	{\cal F}_3= \frac{1}{2^{18} \Delta^4}\sum_{i=0}^6 f_i X^i\,,
	\eea
where the coefficients $f_i$ can be found in Appendix \ref{appF3}. We have also derived ${\cal F}_4$. 
Unfortunately it is too lengthy to be presented here. Anyway we will make it available upon request.
\section{Heavy conformal block}
\label{HeavyCFTblock} 
\subsection{BPZ differential equation for rank $3/2$ conformal block}
Let us find the BPZ differential equation satisfied by
\bea
\label{F32def}
F^{(3/2)}(z,c_1,c_2)=\langle \beta_0|V_{deg}(z) | I^{(3/2)} \rangle\,.
\eea
We know that for the level two degenerate field with charge $-b/2$ one has
\bea
\label{nullvec}
\left(L_{-1}^2+b^2L_{-2}\right)V_{deg}(z)=0\,.
\eea
To obtain the BPZ differential equation let us consider
	\bea
	\langle \beta_0|(L_{-2}V_{deg}(z)) | I^{(3/2)} \rangle&=&
	\frac{1}{2 \pi i}\oint_{{\cal C}_z} d\, w (w-z)^{-1} \langle  V_{\beta_0}(\infty) T(w) V_{deg}(z) | I^{(3/2)} \rangle=
	\\ \nn
	&-& \frac{1}{ 2 \pi i } \oint_{{\cal C}_0} d \, w (w-z)^{-1} \langle V_{\beta_0}(\infty) V_{deg}(z) \sum_{n=-1}^{4}\frac{L_n}{w^{n+2}}| I^{(3/2)} \rangle \,,
	\eea
where ${\cal C}_z$, ${\cal C}_0$ are small contours surrounding $z$ and $0$ respectively. The second equality above 
is due to vanishing of the integral over large contour ${\cal C}_\infty$.
Evaluating the integrals and taking into account (\ref{def_ir32})
\footnote{Since the action $L_{-1} | I^{(2)} \rangle$ is not determined, 
one pushes $L_{-1}$ to the left using $\left[V_{deg}(z), L_{-1}\right]=-V'_{deg}(z)$ 
and $\langle p_0| L_{-1}=0$.
} we obtain
	\bea
	\label{T2op_rank32}
	&\langle \beta_0|(L_{-2}V_{deg}(z)) | I^{(3/2)} \rangle =
	\left(
	\frac{{\cal L}^{(3/2)}_0}{z^2}
	+ \frac{{\cal L}^{(3/2)}_1}{z^3}
	+ \frac{{\cal L}^{(3/2)}_2}{z^4}
	+ \frac{{\cal L}^{(3/2)}_3}{z^5}
	-\frac{1}{z}\frac{\partial}{\partial z}
	\right) F^{(3/2)}\,.
	\eea
Thus (\ref{nullvec}) imposes the differential equation on irregular block (\ref{F32def})
\bea
\label{BPZr32}
& \bigg(
\frac{1}{b^2}\frac{\partial^2}{\partial z^2}-\frac{1}{z}\frac{\partial}{\partial z}
+\frac{1}{z^2}\left(3 \Lambda_3 \frac{\partial}{\partial \Lambda_3}+c_1 \frac{\partial}{\partial c_1}\right)
- \frac{\Lambda_3}{2 c_1z^3} \frac{\partial}{\partial c_1}
- \frac{c_1^2}{z^4}
+ \frac{\Lambda_3}{z^5}
\bigg)F^{(3/2)}=0\,.
\eea
We can eliminate the derivative over $\Lambda_3$ by observing that 
\bea
\label{L_0_l_r2}
\langle \beta_0| L_{0}V_{deg}(z) | I^{(3/2)} \rangle&=&
\Delta_{\beta_0}  F^{(3/2)}\,,
\\
\label{L_0_r_r2}
\langle \beta_0| L_{0}V_{deg}(z) | I^{(3/2)} \rangle &=&
\left(
z \partial_z + \Delta_{deg}
+{\cal L}_0^{(3/2)}
\right) F^{(3/2)}\,,
\eea
where the relation $\left[ L_{0},V_{deg}(z)\right]=\Delta_{deg}V_{deg}(z)+z V'_{deg}(z)$ 
has been used. Hence
\bea
\label{1ord_diffeq_rank2}
\left(
z \partial_z+3 \Lambda_3 \frac{\partial}{\partial \Lambda_3}+c_1 \frac{\partial}{\partial c_1}
+ \Delta_{deg}-\Delta_{\beta_0}
\right) F^{(3/2)}=0\,.
\eea
This relation allows us to eliminate $\Lambda_3$ derivative in (\ref{BPZr32}). Finally we get
\bea
\label{BPZr32NoL3}
\left(
\frac{1}{b^2}\frac{\partial^2}{\partial z^2}-\frac{2}{z}\frac{\partial}{\partial z}
- \frac{\Lambda_3}{2 c_1z^3} \frac{\partial}{\partial c_1}
+\frac{1}{z^2}(\Delta_{\beta_0}-\Delta_{deg})
- \frac{c_1^2}{z^4}
+ \frac{\Lambda_3}{z^5}
\right)F^{(3/2)}\left(z,c_1\right)=0\,.
\eea
\subsection{The deformed  SW curve}
In this section we derive the so-called "heavy" limit of (\ref{BPZr32NoL3}). In our case, 
the heavy limit  on the CFT side is obtained by sending 
 $b \to 0$,  while keeping $b \beta_0$,  $b c_1$ and $b^2 \Lambda_3$ finite.
Consulting section \ref{irr_states}, where the CFT/gauge theory map is described in details and specifying 
$\Omega$-background parameters to $\epsilon_1=b^2$, $\epsilon_2=1$, the reader 
can easily get convinced that the heavy limit in gauge theory side precisely coincides
with Nekrasov-Shatashvili (NS) limit. In the NS limit we keep
\bea
\epsilon_1 \Delta_{\beta_0}=\frac{1}{4}-\hat{k}^2\,,
\quad
\sqrt{\epsilon_1} c_{1}=\hat{c}_{1}\,,
\quad
\epsilon_1\Lambda_{3}=\hat{\Lambda}_{3}\,,
\eea
finite while  sending  $\epsilon_1 \to 0$, as it becomes clear from  (\ref{map_CFT_H1}). In this limit CFT intuition suggests that 
\bea
F^{(3/2)}\sim e^{\frac{1}{\epsilon_1}{\cal F}(\hat{c}_1)}\Psi(z,\hat{c}_1)\,,
\eea
(${\cal F}(\hat{c}_1)$ is the heavy block) with $\Psi(z,\tilde{c}_1)$ satisfying the differential equation
\bea
\label{BPZr32NoL3Heavy}
(\partial_z^2-\hat{\phi}_2(z))\Psi(z,\hat{c}_1)=0\,,
\eea
where
\bea
\label{GswCurve_heavy}
\hat{\phi}_2(z)  =
{\hat{k}^2- \frac{1}{4} \over z^2}  {+}{ 2 \hat{v} \over z^3}
{+}{\hat{c}_1^2\over z^4}
{-}{\hat{\Lambda}_3 \over z^5}\,,
\eea
and
\be
\label{G_Matone_CFT_heavy}
\hat{v}=\frac{\hat{\Lambda}_3}{4\hat{c}_1}\,\partial_{\hat{c}_1}{\cal F}(\hat{c}_1) \,.
\ee
The shift of $k^2$ by  $-1/4$ in $\hat{\phi}_2(z)$  compared with  (\ref{phi2hat}) is due to $\Omega$-background corrections.
(\ref{G_Matone_CFT_heavy}) is  Matone's relation \cite{Matone:1995rx} generalized to the case with $\Omega$-background \cite{Flume:2004rp}.
Our main task in this section is to identify deformed SW curve for 
${\cal H}_1$ theory. Usually DSW curve is derived using explicit combinatorial expressions for gauge 
theory partition functions and chiral correlators. Then analogues of the differential equation 
(\ref{BPZr32NoL3Heavy}) follow from DSW curve equation. In our case we will proceed in opposite way 
starting from (\ref{BPZr32NoL3Heavy}) then recovering the underlying DSW ``curve". 
We start by looking solution for (\ref{BPZr32NoL3Heavy}) of the form
\bea
\Psi(z,\hat{c}_1)=e^{-\frac{\hat{c}_1}{z}}  \sum_{x=\hat{a}+\mathbb{Z}}Y(x) z^{-x}\,.
\eea
Substituting this into (\ref{BPZr32NoL3Heavy}) for $Y(x)$ we get
\bea
\left(x^2+x+\frac{1}{4}-\hat{k}^2\right) Y(x)-2 \left( \hat{c}_1 x+ \hat{v}\right)Y(x-1)
+\hat{\Lambda} _3 Y(x-3)=0\,.
\eea
Let us extend the range of $x$ from the discrete set $\hat{a}+\mathbb{Z}$ to entire 
complex plane $\mathbb{C}$. Then for the function 
\bea
y(x)=\frac{Y(x)}{Y(x-1)}\,.
\eea
we get 
\bea
\label{DSW}
\left(x^2+x+\frac{1}{4}-\hat{k}^2\right) y(x)+\frac{\hat{\Lambda} _3}{y(x-2) y(x-1)}
=2 \left(\hat{c}_1 x+\hat{v}\right)\,,
\eea
which will be referred as the DSW equation for ${\cal H}_1$ theory.
The function $y(x)$ is related to the DSW differential by
\bea
\label{SWdiff}
\lambda_{SW}=-\frac{x}{2 \pi i} \frac{d}{dx}{\rm log }\,y(z)\,.
\eea
The deformed $A$-cycle $\hat{a}$ is given as in (\ref{Acycle}). To get convinced that all these works 
 we have computed $\hat{a}$ up to order $O(\hat{\Lambda}_3^6)$
(some details of calculation can be found in Appendix \ref{sec_Aheavy})
and then inverted the series. Here is the result
	\bea
	\label{Aheavy}
	\frac{\hat{v}}{\hat{c}_1}=
	\sum_{i=0}^{\infty}A_i \left(\frac{{\hat \Lambda}_3}{16 {\hat c}_1^3}\right)^i
	\,.
	\eea
where
\bea
A_0&=&{\hat a}\,,
\qquad
A_1=12 {\hat a}^2-4 {\hat k}^2+1\,,
\qquad
A_2= -4 {\hat a} \left(68 {\hat a}^2-36 {\hat k}^2+19\right) \,,
\\
A_3&=&2 \left(-8 \left(516 {\hat a}^2+71\right) {\hat k}^2+6000 {\hat a}^4+3672 {\hat a}^2+176 {\hat k}^4+131\right)\,,
\\
A_4&=&4 {\hat a} \left(40 {\hat a}^2 \left(3564 {\hat k}^2-4681\right)-171024 {\hat a}^4+72 {\hat k}^2 \left(821-202 {\hat k}^2\right)-22709\right)\,,
\\
A_5&=&16 \bigg(-23520 {\hat a}^4 \left(116 {\hat k}^2-207\right)+6 {\hat a}^2 \left(76400 {\hat k}^4-381400 {\hat k}^2+217663\right)
\\ \nn
&&\qquad \qquad \qquad
+2801568 {\hat a}^6-7360 {\hat k}^6+72944 {\hat k}^4-143572 {\hat k}^2+31449\bigg)\,.
\eea
We checked that above expression completely agrees with holomorphic anomaly result.
Additionally, we performed exact WKB analysis (see e.g. \cite{Fioravanti:2019vxi,Fucito:2023plp}) up to order $8$ which once again confirms this result.
We end this section by emphasizing the advantage of DSW method compared to WKB, since the former 
is exact in $\Omega$-background parameter $\epsilon_2$ (playing the role of Plank's constant in WKB setting).  
\acknowledgments
The research of R.P. was  supported by the Armenian SCS  grants 21AG-1C060 and 24WS-1C031.
 Similarly, H.P.'s work received support from   Armenian SCS  grant  21AG-1C062 and 24WS-1C031.

\begin{appendix}
\section{Irregular states}
\subsection{Integer rank irregular states}
\label{IntegerRankIrregularStates}
The rank $n$ irregular states  $ |I^{(n)}\rangle $  in 2d Liouville conformal field theory, which 
depend on two sets of parameters ${\bf c}=\{c_0,\ldots ,c_n\}$ and ${\bm \beta}
=\{\beta_0, \ldots,\beta_{n-1}\}$ are defined by
\cite{Gaiotto:2012sf}
\bea
\label{GaiottoDef}
L_k |I^{(n)}( {\bf c},{\bm \beta} )\rangle  &=& {\cal L}^{(n)}_k|I^{(n)}( {\bf c},{\bm \beta} )\rangle \, ,\quad k=0,\ldots n-1 \nn\\
L_k |I^{(n)}( {\bf c},{\bm \beta} )\rangle  &=& \Lambda^{(n)}_k|I^{(n)}( {\bf c},{\bm \beta} )\rangle \, ,\quad k=n,\ldots 2n\nn\\
L_k |I^{(n)}( {\bf c},{\bm \beta} )\rangle  &=& 0  \, ,\quad\qquad\qquad ~~~~k> 2n
\label{lneq}
\eea
with
\begin{small}
	\bea
	{\cal L}^{(n)}_k & =&  (k+1)Qc_k -\sum_{\ell=0}^{k} c_\ell \, c_{k-\ell}
	+\sum_{\ell=1}^{n-k}\ell c_{k+\ell}{\partial \over \partial c_{\ell} }\,,  \quad  k=0,\ldots , n-1 \nn\\
	\Lambda^{(n)}_k   &=& (n+1)Qc_n\delta_{k,n} -\sum_{\ell=k-n}^{n}c_{\ell}  c_{k-\ell} \,, 
	\qquad  k=n,\ldots , 2n \label{lambdak}\,.
	\eea
\end{small}
Notice that the differential operators ${\cal L}^{(n)}_k$ are constructed in 
such a way, that above relations are compatible with Virasoro algebra commutation 
rules. Namely, the form of this operators closely resembles the famous Feigin-Fuchs 
representation of Virasoro algebra in terms of free boson oscillators $c_k$. The meaning 
of the second set of parameters $\bm \beta $ is more subtle. These are remnants of internal 
Liouville momenta specifying successive OPE structure of those $n$ primary fields, 
which in colliding limit create the irregular state under discussion (see \cite{Gaiotto:2012sf} 
for details).    

The state  $|I^{(n)}({\bf c},{\bm \beta} )\rangle $ can be expanded in $c_n$ power series
\be
|I^{(n)}( {\bf c},{\bm \beta} )\rangle  =
f({\bf c}, \beta_{n-1} )  \sum_{k=0}^\infty c_n^k   |I^{(n-1)}_k( \tilde {\bf c},\tilde {\bm\beta} ) \rangle\,,
\ee
where 
\be
\tilde{\bf c} =(\beta_{n-1}  ,c_1,\ldots c_{n-1} ) \qquad , \qquad   \tilde{\beta}=(\beta_0,\beta_1\ldots \beta_{n-2})\,.
\ee
Here $  |I^{(n-1)}_k( \tilde {\bf c},\tilde {\bm\beta} ) \rangle $ is a level $k$ generalized descendant of rank 
$n-1$ irregular state  $ |I^{(n-1)}( \tilde {\bf c},\tilde {\bm\beta} )\rangle $ obtained by acting with Virasoro generators and derivatives with respect to $c_1,\ldots c_{n-1}$. It is argued in \cite{Gaiotto:2012sf} that 
after specifying the prefactor $f({\bf c}, \beta_{n-1} )$ appropriately these descendants can be determined order by order uniquely imposing equations (\ref{lneq}).

\subsection{Irregular states of (Poincar\'{e}) rank $3/2$}
\label{ZH1LambdaExpansion}
The rank $3/2$ irregular state admits a small $\Lambda_3$ expansion
\be
\label{rank32_vs_rank2_expansion}
|I^{(3/2)}(c_1,\Lambda_3;\beta_0,c_0)\rangle  =
f(c_0,c_1,\Lambda_3 )  \sum_{k=0}^\infty \Lambda_3^k   |I^{(1)}_k(c_0,c_1;\beta_0 ) \rangle\,.
\ee
The leading term $|I^{(1)}_0(c_0,c_1;\beta_0 ) \rangle$ is just the rank $1$ irregular state, while 
the generalized descendants are some linear combinations of monomials\footnote{Given a partition 
	$Y=1^{n_1}2^{n_2}3^{n_3}\cdots$, by definition $L_{-Y}=\cdots L_{-3}^{n_3}L_{-2}^{n_2}L_{-1}^{n_1}$.}
\bea 
\label{desst}
L_{-Y}c_1^{-r_1}\partial_{c_1}^{m_1}|I^{(1)}_0(c_0,c_1;\beta_0 ) \rangle\,,
\eea  
where $n=|Y|$, $r_{1}$, $m_{1}$ are non-negative integers, subject to constraint 
\bea 
3k=n+m_1+r_1\,.
\eea
In addition, the maximum power of  $c_1 $
at a given level $k$ is constrained by the condition $r_1\le 3k$. 
Though the number of permitted terms increases significantly with the level, it remains finite for any given 
$k$. 

Following the line of reasoning presented in \cite{Poghosyan:2023zvy} for the rank $5/2$, it is possible to argue 
that the appropriate prefactor for the rank $3/2$ case is
\bea
\label{factor_f}
f(c_0,c_1,\Lambda_3 )=c_1^{\rho_1}  \Lambda _3^{\rho_3}e^{-\frac{4 c_1^3 (Q-c_0)}{3 \Lambda _3}}\,.
\eea
Here  $\rho_1$ and $\rho_2$ are given by equations (\ref{rho1}) and (\ref{rho3}) below. 

We can recover the  expansion coefficients of  (\ref{rank32_vs_rank2_expansion})  order by order in $\Lambda_3$
with the following procedure.
From (\ref{def_ir32}), (\ref{GaiottoDef}) and (\ref{rank32_vs_rank2_expansion}) for any $m=0,1,\ldots $ we have
\bea
\label{recRelLambda}
& \sum_{k=0}^\infty \left( {\cal L}^{(3/2)}_m \left(f(c_1,\Lambda_3 )  \Lambda_3^k   |I^{(1)}_k \rangle\right)-
f(c_1,\Lambda_3 )  \Lambda_3^k   {\cal L}^{(1)}_m |I^{(1)}_k \rangle \right)
=0\,.
\eea
At level $0$ we get
\bea
{\cal L}^{(3/2)}_0(f(c_1,\Lambda_3 )|I^{(1)}_0 \rangle)=
f(c_1,\Lambda_3 ){\cal L}^{(1)}_0|I^{(1)}_0 \rangle \,.
\eea
Using (\ref{lambdak}), (\ref{ln_32}) and (\ref{factor_f}) we  obtain
\bea
\label{rho3}
\rho_3=\frac{1}{3} \left(c_0( Q-c_0)-\rho_1\right)\,.
\eea
To get the next term in expansion (\ref{rank32_vs_rank2_expansion}) we notice that 
due to (\ref{desst})
\bea
\label{FirstDesIr1} 
|I^{(1)}_1 \rangle=\left(
\frac{s_1}{c_1^3}+\frac{ s_2}{c_1^2}L_{-1}+\frac{s_3}{c_1^2}\partial _{c_1}
+\frac{s_4}{c_1}\partial_{c_1}^2
+\frac{s_5}{c_1}L_{-1}\partial _{c_1}
\right)
|I^{(1)}_0 \rangle
\eea
with some coefficients $s_{1,\ldots,5}$.
Inserting this into  (\ref{recRelLambda}) with $m=1,2$, we get  nontrivial equations determining the coefficients
$s_{2,3,4,5}$ and $\rho_1$:
\bea
\label{rho1} 
\rho_1=\left(3 Q-2 c_0\right) \left(Q-c_0\right)\,,
\eea
\bea
& s_2= -\frac{1}{4}\,,
\quad
s_3= \frac{3 (c_0-Q)}{4}\,,
\quad
s_4= 0\,,
\quad
s_5= 0\,.
\eea
The remaining coefficient $s_1$ is determined from the second order in $\Lambda_3$ computation with result
\bea
s_1=-\frac{1}{144} \left(Q-c_0\right) \left(-300 c_0 Q+96 c_0^2+5 c+204 Q^2-11\right)\,.
\eea

To proceed we need to calculate the amplitude $\langle \Delta_{\beta_0}  |I_0^{(1)}\rangle $.
Consider $\langle \Delta_{\beta_0} |L_0|I_0^{(1)}\rangle $ taking into account that $L_0$ acts diagonally on the 
left, while on the right it acts by the differential operator
${\cal L}^{(1)}_{0}$ defined in (\ref{lambdak}). This argument immediately leads to the differential 
equation
\bea
c_0(Q-c_0) -\Delta_{\beta_0} +c_1 \frac{\partial}{\partial c_1} \log \langle \Delta_{\beta_0} |I^{(1)}\rangle =0 \,,
\eea    
which up to a $c_1$ independent factor gives
\bea 
\label{vac_ir2}
\langle \Delta_{\beta_0}  |I_0^{(1)}\rangle =c_1^{\Delta_{\beta_0}-c_0(Q-c_0)} \,.
\eea  
Plugging (\ref{rank32_vs_rank2_expansion}) into (\ref{ZH1_def}) and taking into account 
(\ref{factor_f}), (\ref{FirstDesIr1}) and, (\ref{vac_ir2}) one finds $Z_{{\cal H}_1}=Z_{{\cal H}_1,\rm tree}Z_{{\cal H}_1,\rm inst}$ with
\bea
\label{Zh1tree}
Z_{ {\cal H}_1 \rm tree} &=& c_1^{\Delta_{\beta_0}-c_0(Q-c_0)+\rho_1} \Lambda_3^{\rho_3}
e^{-\frac{4 c_1^3 (Q-c_0)}{3 \Lambda _3}}\,,\\
\label{Zh1inst}
Z_{ {\cal H}_1 \rm inst} &=&
1{+}\frac{ \left(c_0-Q\right) \left(34 \left(c_0-Q\right){}^2+18 \Delta_{\beta_0}+5 Q^2-1\right)}{24 c_1^3}\Lambda _3 {+}\ldots .
\eea
Higher order  terms in  $\Lambda_3$ can be found in \cite{Hamachika:2024efr}. The notations there are the same 
besides that ${\cal Z}_{{\cal H}_1}$ is denoted as ${\cal Z}_{(A_1,D_3)}$ and $c_0$ by $\beta_1$.
\section{Expression for ${\cal F}_3$}
Here we give explicit expressions for the coefficients $f_i$ in (\ref{F3_fi_rel}).
\label{appF3}
	\begin{footnotesize}
	\bea
	f_6=3^{10} 5  w^4 \,,
	\qquad
	f_5=3^7 5 w^2  \left(\left(Q^2-16\right) w \Delta '+20 \Delta  w'\right)\,,
	\eea
	\bea 
	f_4=81\bigg(324 g_2 w ^4+w ^2 \left(\left(2 Q^4-107 Q^2+633\right) \Delta '^2+12 \left(5 Q^2-26\right) \Delta  \Delta ''\right)+192 \Delta ^2 w '^2
	\\ \nn
	+4 \Delta w  \left(\left(16 Q^2-207\right) \Delta ' w '+84 \Delta  w ''\right)
	\bigg)\,,
	\eea
\end{footnotesize}
\begin{tiny}
	\bea
	f_3=w \left(12 \Delta  \left(12 Q^4-425 Q^2+944\right) \Delta ' \Delta ''+\left(Q^6-162 Q^4+3875 Q^2-9294\right) \Delta '^3+144 \Delta ^2 \Delta ^{(3)} \left(10 Q^2-19\right)\right)
	\\ \nn
	+
	12 \Delta  \left( w' \left(\left(3 \left(Q^2-41\right) Q^2+623\right) \Delta '^2+24 \Delta  \left(3 Q^2-16\right) \Delta ''\right)+36 \Delta  \left(\left(Q^2-9\right) \Delta ' w''+4 \Delta  w^{(3)}\right) \right)
	\\ \nn
	+27 w^3 \left(156 \Delta  g_2'+g_2 \left(31 Q^2-578\right) \Delta '\right)+17820 \Delta  g_2 w^2 w'\,,
	\eea
	\bea 
	f_2=\frac{36}{5} \Delta  w \left(g_2 \left(\left(-137 Q^4+403 Q^2-934\right) \Delta ' w'+12 \Delta  \left(7 Q^4-18 Q^2+29\right) w''\right)+12 \Delta  \left(7 Q^4-18 Q^2+34\right) g_2' w'\right)
	\\ \nn
	+1296 \Delta ^2 g_2 w'^2+729 g_2^2 w^4-\frac{1}{9} \left(Q^2-1\right) \bigg[-12 \Delta  \left(Q^4-88 Q^2+1485\right) \Delta '^2 \Delta ''+11 \left(Q^4-57 Q^2+782\right) \Delta '^4
	\\ \nn
	+144 \Delta ^2 \Delta ^{(3)} \left(43-2 Q^2\right) \Delta '-144 \Delta ^2 \left(12 \Delta  \Delta ^{(4)}+\left(Q^2-33\right) \Delta ''^2\right)\bigg]
	-\frac{648}{5} \left(39 Q^4-96 Q^2+43\right) w^3 \left(2 g_3 \Delta '-\Delta  g_3'\right)
	\\ \nn
	+\frac{3}{5} w^2 \bigg[g_2 \left(\left(1292 Q^4-4353 Q^2+8644\right) \Delta'^2-12 \Delta  \left(53 Q^4-182 Q^2+311\right) \Delta ''\right)
	\\ \nn
	+12 \Delta  \left(60 \Delta  g_2''+\left(-53 Q^4+157 Q^2-346\right) \Delta ' g_2'+36 g_3 \left(39 Q^4-96 Q^2+43\right) w'\right)\bigg]\,,
	\eea
\end{tiny}
{\fontsize{4.5}{8}	
	\bea
	45 f_1=
	g_2 w \left(12 \Delta  \left(-53 Q^6+3988 Q^4-11125 Q^2+5912\right) \Delta ' \Delta ''+\left(1277 Q^6-49610 Q^4+131467 Q^2-68014\right) \Delta '^3-144 \Delta ^2 \Delta ^{(3)} \left(53 Q^4-162 Q^2+91\right)\right)
	\\ \nn
	+12 \Delta  \left(12 \Delta  \left(\left(7 Q^6-292 Q^4+725 Q^2-358\right) \Delta ' g_2' w'+12 \Delta  \left(7 Q^4-18 Q^2+9\right) \left(2 g_2' w''+g_2'' w'\right)+36 g_3 \left(39 Q^4-96 Q^2+43\right) w'^2\right)\right)
	\\ \nn
	+12 \Delta  w \left(24 \Delta  \left(\left(-53 Q^4+147 Q^2-76\right) \Delta '' g_2'+36 \left(39 Q^4-96 Q^2+43\right) g_3' w'\right)+\left(-53 Q^6+2691 Q^4-7007 Q^2+3523\right) \Delta '^2 g_2'\right)
	\\ \nn
	-216 w^2 \left(\left(39 Q^4-96 Q^2+43\right) \left(2 g_3 \left(12 \Delta  \Delta ''+\left(Q^2-36\right) \Delta '^2\right)+\Delta  \left(-12 \Delta  g_3''-\left(Q^2-48\right) \Delta ' g_3'\right)\right)-90 \Delta  g_2^2 w'\right)
	\\ \nn
	+144 \Delta ^2 g_2 \left(\left(7 Q^6-239 Q^4+583 Q^2-287\right) \Delta ' w''+2 \left(6 \Delta  \left(7 Q^4-18 Q^2+9\right) w^{(3)}-5 \left(19 Q^4-51 Q^2+26\right) \Delta '' w'\right)\right)
	\\ \nn
	+12 \Delta  w \left(36 g_3 \left(39 Q^4-96 Q^2+43\right) \left(12 \Delta  w''+\left(Q^2-48\right) \Delta ' w'\right)-12 \Delta  \left(53 Q^4-142 Q^2+71\right) \Delta ' g_2''\right)
	\\ \nn
	+810 g_2 w^3 \left(12 \Delta  g_2'+g_2 \left(Q^2-23\right) \Delta '\right)+12 \Delta  g_2 \left(-137 Q^6+4923 Q^4-12299 Q^2+6115\right) \Delta '^2 w'\,.
	\eea
}
The $f_0$ term is derived from the gap condition (\ref{gap_condition})
{\fontsize{4.5}{8}\bea
\frac{f_0}{{\tilde \Lambda}_3^4}=-\frac{36864}{35} {\tilde g}_3^4 \left(243 {\tilde g}_3^2-72 \left(3 {\tilde g}_2+4\right) {\tilde g}_3+8 {\tilde g}_2 \left(9 {\tilde g}_2+2\right)\right) \left(923113 Q^6-4126428 Q^4+5023827 Q^2-1491431\right)
\\ \nn
-\frac{4096}{45}  {\tilde \Delta} ^3 \left(1133 Q^6-5478 Q^4+7124 Q^2-2240\right)
+\frac{4096}{2835}  {\tilde \Delta} ^2 \bigg[-243 {\tilde g}_3^2 \left(417803 Q^6-1896684 Q^4+2348446 Q^2-710663\right)
\\ \nn
+36{\tilde  g}_3 \left(3 {\tilde g}_2 \left(436459 Q^6-1981551 Q^4+2457011 Q^2-745826\right)+4 \left(773583 Q^6-3483033 Q^4+4282255 Q^2-1289272\right)\right)
\\ \nn
+4 {\tilde g}_2 \left(-9 {\tilde g}_2 \left(134186 Q^6-608626 Q^4+755418 Q^2-230145\right)-403992 Q^6+1826712 Q^4-2257516 Q^2+685400\right)\bigg]
\\ \nn
+\frac{4096}{105} {\tilde \Delta}  {\tilde g}_3^2 \bigg[
8 {\tilde g}_2 \left(3 {\tilde g}_2 \left(3629779-2191983 Q^6+9859098 Q^4-12097861 Q^2\right)+2846622-1740460 Q^6+7801212 Q^4-9534632 Q^2\right)
\\ \nn
+36 {\tilde g}_3 \left(3 {\tilde g}_2 \left(2089349 Q^6-9390987 Q^4+11508811 Q^2-3445420\right)+4 \left(2575843 Q^6-11544711 Q^4+14105617 Q^2-4208828\right)\right)
\\ \nn
-243 {\tilde g}_3^2 \left(1306849 Q^6-5873310 Q^4+7194337 Q^2-2151334\right)
\bigg]\,.
\eea}
Here we have used the notations
\bea
{\tilde g}_2 = \frac{g_2}{c_1^4}\,,
\quad
{\tilde g}_3 = \frac{g_3}{c_1^6}\,,
\quad
{\tilde \Delta} = \frac{\Delta}{c_1^{12}}\,,
\quad
{\tilde \Lambda}_3=c_1^9 \Lambda_3\,.
\eea
\section{The $A$-period from DSW}
\label{sec_Aheavy}
In this appendix we will derive the $A$-period (\ref{Aheavy}) using DSW curve equation (\ref{DSW}).
Inserting the expansion
\bea
y(x)=y_0(x)+y_1(x)\Lambda_3+y_2(x)\Lambda_3^2+\dots+O(\Lambda_3^{n+1})
\eea
into  (\ref{DSW}) we get
\begin{footnotesize}
	\bea
	&-&2 \left(\tilde{c}_1 x+\tilde{v}\right)+\left(\frac{1}{4}-\hat{k}^2+x^2+x\right) y_0(x)
	+\Lambda _3 \left(\left(\frac{1}{4}-\hat{k}^2+x^2+x\right) y_1(x)
	+\frac{1}{y_0(x-2) y_0(x-1)}\right)
	\\ \nn
	&+&\Lambda _3^2 \left(\left(\frac{1}{4}-\hat{k}^2+x^2+x\right) y_2(x)-\frac{y_0(x-1) y_1(x-2)
		+y_0(x-2) y_1(x-1)}{y_0(x-2)^2 y_0(x-1)^2}\right)+\dots+O(\Lambda_3^{n+1})=0\,.
	\eea
\end{footnotesize}
This implies
\bea
&&y_0(x)=\frac{2 \left(\hat{c}_1 x+\hat{v}\right)}{\frac{1}{4}-\hat{k}^2+x^2+x}\,,
\\
&&y_1(x)=-\frac{(\frac{1}{4}-\hat{k}^2+(x-1) x) \left(\frac{1}{4}-\hat{k}^2+x^2-3 x+2\right)}{4 \left(\frac{1}{4}-\hat{k}^2+x^2+x\right) \left(\hat{c}_1 (x-2)+\hat{v}\right) \left(\hat{c}_1 (x-1)+\hat{v}\right)}\,,
\eea
\begin{footnotesize}
	\bea
	y_2(x)=
	-\frac{(\frac{1}{4}-\hat{k}^2+(x-1) x) \left(\frac{1}{4}-\hat{k}^2+x^2-5 x+6\right) \left(\frac{1}{4}-\hat{k}^2+x^2-3 x+2\right)}
	{32 \left(\frac{1}{4}-\hat{k}^2+x^2+x\right) \left(\hat{c}_1 (x-4)+\hat{v}\right) \left(\hat{c}_1 (x-3)+\hat{v}\right) \left(\hat{c}_1 (x-2)+\hat{v}\right){}^2 \left(\hat{c}_1 (x-1)+\hat{v}\right)^2}
	\\ \nn
	 \times \left(\hat{c}_1 (2 x-5) \left(\frac{1}{4}-\hat{k}^2+x^2-5 x+4\right)+2 \hat{v} (\frac{1}{4}-\hat{k}^2+(x-5) x+7)\right)\,,
	 \\ \nn
	 \vdots
	  \qquad \qquad \qquad \qquad  \qquad \qquad \qquad \qquad \qquad
	\eea
\end{footnotesize}
From (\ref{SWdiff}) and (\ref{Acycle}), taking into account that 
the contour ${\cal C}_A$ includes only the poles
\bea
-\frac{\hat{v}}{\hat{c}_1}\,,
\quad
-\frac{\hat{v}}{\hat{c}_1}+1\,,
\quad
-\frac{\hat{v}}{\hat{c}_1}+2\,,
\quad
\dots \,,
-\frac{\hat{v}}{\hat{c}_1}+2 n\,,
\eea
we have calculated  $\hat{a}$, inverting which we got  equation (\ref{Aheavy}).
Actual computation were done up to order $n=5$.
\end{appendix}

\bibliographystyle{JHEP}

\begin{thebibliography}{10}
	
	\bibitem{Seiberg:1994rs}
	N.~Seiberg and E.~Witten, {\it {Electric - magnetic duality, monopole
			condensation, and confinement in N=2 supersymmetric Yang-Mills theory}},
	{\em Nucl. Phys. B} {\bf 426} (1994) 19--52,
	[\href{http://arxiv.org/abs/hep-th/9407087}{{\tt hep-th/9407087}}]. [Erratum:
	Nucl.Phys.B 430, 485--486 (1994)].
	
	\bibitem{Seiberg:1994aj}
	N.~Seiberg and E.~Witten, {\it {Monopoles, duality and chiral symmetry breaking
			in N=2 supersymmetric QCD}},  {\em Nucl. Phys. B} {\bf 431} (1994) 484--550,
	[\href{http://arxiv.org/abs/hep-th/9408099}{{\tt hep-th/9408099}}].
	
	\bibitem{Bellisai:2000bc}
	D.~Bellisai, F.~Fucito, A.~Tanzini, and G.~Travaglini, {\it {Instanton
			calculus, topological field theories and N=2 superYang-Mills theories}},
	{\em JHEP} {\bf 07} (2000) 017,
	[\href{http://arxiv.org/abs/hep-th/0003272}{{\tt hep-th/0003272}}].
	
	\bibitem{Flume:2001kb}
	R.~Flume, R.~Poghossian, and H.~Storch, {\it {The Seiberg-Witten prepotential
			and the Euler class of the reduced moduli space of instantons}},  {\em Mod.
		Phys. Lett. A} {\bf 17} (2002) 327--340,
	[\href{http://arxiv.org/abs/hep-th/0112211}{{\tt hep-th/0112211}}].
	
	\bibitem{Flume:2001nc}
	R.~Flume, R.~Poghossian, and H.~Storch, {\it {The Coefficients of the
			Seiberg-Witten prepotential as intersection numbers(?)}},
	\href{http://arxiv.org/abs/hep-th/0110240}{{\tt hep-th/0110240}}.
	
	\bibitem{Dorey:2002ik}
	N.~Dorey, T.~J. Hollowood, V.~V. Khoze, and M.~P. Mattis, {\it {The Calculus of
			many instantons}},  {\em Phys. Rept.} {\bf 371} (2002) 231--459,
	[\href{http://arxiv.org/abs/hep-th/0206063}{{\tt hep-th/0206063}}].
	
	\bibitem{Nekrasov:2002qd}
	N.~A. Nekrasov, {\it {Seiberg-Witten prepotential from instanton counting}},
	{\em Adv. Theor. Math. Phys.} {\bf 7} (2003), no.~5 831--864,
	[\href{http://arxiv.org/abs/hep-th/0206161}{{\tt hep-th/0206161}}].
	
	\bibitem{Flume:2002az}
	R.~Flume and R.~Poghossian, {\it {An Algorithm for the microscopic evaluation
			of the coefficients of the Seiberg-Witten prepotential}},  {\em Int. J. Mod.
		Phys. A} {\bf 18} (2003) 2541,
	[\href{http://arxiv.org/abs/hep-th/0208176}{{\tt hep-th/0208176}}].
	
	\bibitem{Bruzzo:2002xf}
	U.~Bruzzo, F.~Fucito, J.~F. Morales, and A.~Tanzini, {\it {Multiinstanton
			calculus and equivariant cohomology}},  {\em JHEP} {\bf 05} (2003) 054,
	[\href{http://arxiv.org/abs/hep-th/0211108}{{\tt hep-th/0211108}}].
	
	\bibitem{Nekrasov:2003rj}
	N.~Nekrasov and A.~Okounkov, {\it {Seiberg-Witten theory and random
			partitions}},  {\em Prog. Math.} {\bf 244} (2006) 525--596,
	[\href{http://arxiv.org/abs/hep-th/0306238}{{\tt hep-th/0306238}}].
	
	\bibitem{Flume:2004rp}
	R.~Flume, F.~Fucito, J.~F. Morales, and R.~Poghossian, {\it {Matone's relation
			in the presence of gravitational couplings}},  {\em JHEP} {\bf 04} (2004)
	008, [\href{http://arxiv.org/abs/hep-th/0403057}{{\tt hep-th/0403057}}].
	
	\bibitem{Gaiotto:2009we}
	D.~Gaiotto, {\it {N=2 dualities}},  {\em JHEP} {\bf 08} (2012) 034,
	[\href{http://arxiv.org/abs/0904.2715}{{\tt arXiv:0904.2715}}].
	
	\bibitem{Alday:2009aq}
	L.~F. Alday, D.~Gaiotto, and Y.~Tachikawa, {\it {Liouville Correlation
			Functions from Four-dimensional Gauge Theories}},  {\em Lett. Math. Phys.}
	{\bf 91} (2010) 167--197, [\href{http://arxiv.org/abs/0906.3219}{{\tt
			arXiv:0906.3219}}].
	
	\bibitem{Poghossian:2009mk}
	R.~Poghossian, {\it {Recursion relations in CFT and N=2 SYM theory}},  {\em
		JHEP} {\bf 12} (2009) 038, [\href{http://arxiv.org/abs/0909.3412}{{\tt
			arXiv:0909.3412}}].
	
	\bibitem{Argyres:1995jj}
	P.~C. Argyres and M.~R. Douglas, {\it {New phenomena in SU(3) supersymmetric
			gauge theory}},  {\em Nucl. Phys. B} {\bf 448} (1995) 93--126,
	[\href{http://arxiv.org/abs/hep-th/9505062}{{\tt hep-th/9505062}}].
	
	\bibitem{Gaiotto:2009ma}
	D.~Gaiotto, {\it {Asymptotically free $\mathcal{N} = 2$ theories and irregular
			conformal blocks}},  {\em J. Phys. Conf. Ser.} {\bf 462} (2013), no.~1
	012014, [\href{http://arxiv.org/abs/0908.0307}{{\tt arXiv:0908.0307}}].
	
	\bibitem{Gaiotto:2012sf}
	D.~Gaiotto and J.~Teschner, {\it {Irregular singularities in Liouville theory
			and Argyres-Douglas type gauge theories, I}},  {\em JHEP} {\bf 12} (2012)
	050, [\href{http://arxiv.org/abs/1203.1052}{{\tt arXiv:1203.1052}}].
	
	\bibitem{Nishinaka:2012kn}
	T.~Nishinaka and C.~Rim, {\it {Matrix models for irregular conformal blocks and
			Argyres-Douglas theories}},  {\em JHEP} {\bf 10} (2012) 138,
	[\href{http://arxiv.org/abs/1207.4480}{{\tt arXiv:1207.4480}}].
	
	\bibitem{Kanno:2013vi}
	H.~Kanno, K.~Maruyoshi, S.~Shiba, and M.~Taki, {\it {$\mathcal{W}_3$ irregular
			states and isolated $\mathcal{N}$ = 2 superconformal field theories}},  {\em
		JHEP} {\bf 03} (2013) 147, [\href{http://arxiv.org/abs/1301.0721}{{\tt
			arXiv:1301.0721}}].
	
	\bibitem{Bonelli:2016qwg}
	G.~Bonelli, O.~Lisovyy, K.~Maruyoshi, A.~Sciarappa, and A.~Tanzini, {\it {On
			Painlev\'e/gauge theory correspondence}},  {\em Lett. Matth. Phys.} {\bf 107}
	(2017), no.~12 2359--2413, [\href{http://arxiv.org/abs/1612.06235}{{\tt
			arXiv:1612.06235}}].
	
	\bibitem{Nishinaka:2019nuy}
	T.~Nishinaka and T.~Uetoko, {\it {Argyres-Douglas theories and Liouville
			Irregular States}},  {\em JHEP} {\bf 09} (2019) 104,
	[\href{http://arxiv.org/abs/1905.03795}{{\tt arXiv:1905.03795}}].
	
	\bibitem{Kimura:2020krd}
	T.~Kimura, T.~Nishinaka, Y.~Sugawara, and T.~Uetoko, {\it {Argyres-Douglas
			theories, S-duality and AGT correspondence}},  {\em JHEP} {\bf 04} (2021)
	205, [\href{http://arxiv.org/abs/2012.14099}{{\tt arXiv:2012.14099}}].
	
	\bibitem{Bonelli:2021uvf}
	G.~Bonelli, C.~Iossa, D.~P. Lichtig, and A.~Tanzini, {\it {Exact solution of
			Kerr black hole perturbations via CFT2 and instanton counting: Greybody
			factor, quasinormal modes, and Love numbers}},  {\em Phys. Rev. D} {\bf 105}
	(2022), no.~4 044047, [\href{http://arxiv.org/abs/2105.04483}{{\tt
			arXiv:2105.04483}}].
	
	\bibitem{Bonelli:2022ten}
	G.~Bonelli, C.~Iossa, D.~Panea~Lichtig, and A.~Tanzini, {\it {Irregular
			Liouville Correlators and Connection Formulae for Heun Functions}},  {\em
		Commun. Math. Phys.} {\bf 397} (2023), no.~2 635--727,
	[\href{http://arxiv.org/abs/2201.04491}{{\tt arXiv:2201.04491}}].
	
	\bibitem{Kimura:2022yua}
	T.~Kimura and T.~Nishinaka, {\it {On the Nekrasov partition function of gauged
			Argyres-Douglas theories}},  {\em JHEP} {\bf 01} (2023) 030,
	[\href{http://arxiv.org/abs/2206.10937}{{\tt arXiv:2206.10937}}].
	
	\bibitem{Consoli:2022eey}
	D.~Consoli, F.~Fucito, J.~F. Morales, and R.~Poghossian, {\it {CFT description
			of BH\textquoteright{}s and ECO\textquoteright{}s: QNMs, superradiance,
			echoes and tidal responses}},  {\em JHEP} {\bf 12} (2022) 115,
	[\href{http://arxiv.org/abs/2206.09437}{{\tt arXiv:2206.09437}}].
	
	\bibitem{Fucito:2023plp}
	F.~Fucito, J.~F. Morales, and R.~Poghossian, {\it {On irregular states and
			Argyres-Douglas theories}},  \href{http://arxiv.org/abs/2306.05127}{{\tt
			arXiv:2306.05127}}.
	
	\bibitem{Poghosyan:2023zvy}
	H.~Poghosyan and R.~Poghossian, {\it {A note on rank 5/2 Liouville irregular
			block, Painlev\'e I and the $ \mathcal{H} _{0}$ Argyres-Douglas theory}},
	{\em JHEP} {\bf 11} (2023) 198, [\href{http://arxiv.org/abs/2308.09623}{{\tt
			arXiv:2308.09623}}].
	
	\bibitem{Hamachika:2024efr}
	R.~Hamachika, T.~Nakanishi, T.~Nishinaka, and S.~Tanigawa, {\it {Liouville
			irregular states of half-integer ranks}},  {\em JHEP} {\bf 06} (2024) 112,
	[\href{http://arxiv.org/abs/2401.14662}{{\tt arXiv:2401.14662}}].
	
	\bibitem{Huang:2006si}
	M.-x. Huang and A.~Klemm, {\it {Holomorphic Anomaly in Gauge Theories and
			Matrix Models}},  {\em JHEP} {\bf 09} (2007) 054,
	[\href{http://arxiv.org/abs/hep-th/0605195}{{\tt hep-th/0605195}}].
	
	\bibitem{Huang:2009md}
	M.-x. Huang and A.~Klemm, {\it {Holomorphicity and Modularity in Seiberg-Witten
			Theories with Matter}},  {\em JHEP} {\bf 07} (2010) 083,
	[\href{http://arxiv.org/abs/0902.1325}{{\tt arXiv:0902.1325}}].
	
	\bibitem{Huang:2011qx}
	M.-x. Huang, A.-K. Kashani-Poor, and A.~Klemm, {\it {The $\Omega$ deformed
			B-model for rigid $\mathcal{N}=2$ theories}},  {\em Annales Henri Poincare}
	{\bf 14} (2013) 425--497, [\href{http://arxiv.org/abs/1109.5728}{{\tt
			arXiv:1109.5728}}].
	
	\bibitem{Huang:2013eja}
	M.-x. Huang, {\it {Modular anomaly from holomorphic anomaly in mass deformed
			$\mathcal{N}=2$ superconformal field theories}},  {\em Phys. Rev. D} {\bf 87}
	(2013), no.~10 105010, [\href{http://arxiv.org/abs/1302.6095}{{\tt
			arXiv:1302.6095}}].
	
	\bibitem{Poghossian:2010pn}
	R.~Poghossian, {\it {Deforming SW curve}},  {\em JHEP} {\bf 04} (2011) 033,
	[\href{http://arxiv.org/abs/1006.4822}{{\tt arXiv:1006.4822}}].
	
	\bibitem{Fucito:2011pn}
	F.~Fucito, J.~F. Morales, D.~R. Pacifici, and R.~Poghossian, {\it {Gauge
			theories on $\Omega$-backgrounds from non commutative Seiberg-Witten
			curves}},  {\em JHEP} {\bf 05} (2011) 098,
	[\href{http://arxiv.org/abs/1103.4495}{{\tt arXiv:1103.4495}}].
	
	\bibitem{Nekrasov:2013xda}
	N.~Nekrasov, V.~Pestun, and S.~Shatashvili, {\it {Quantum geometry and quiver
			gauge theories}},  {\em Commun. Math. Phys.} {\bf 357} (2018), no.~2
	519--567, [\href{http://arxiv.org/abs/1312.6689}{{\tt arXiv:1312.6689}}].
	
	\bibitem{Poghosyan:2020zzg}
	H.~Poghosyan, {\it {Recursion relation for instanton counting for SU(2) $
			\mathcal{N} $ = 2 SYM in NS limit of $\Omega$ background}},  {\em JHEP} {\bf
		05} (2021) 088, [\href{http://arxiv.org/abs/2010.08498}{{\tt
			arXiv:2010.08498}}].
	
	\bibitem{Nekrasov:2009rc}
	N.~A. Nekrasov and S.~L. Shatashvili, {\it {Quantization of Integrable Systems
			and Four Dimensional Gauge Theories}},  in {\em {16th International Congress
			on Mathematical Physics}}, pp.~265--289, 8, 2009.
	\newblock \href{http://arxiv.org/abs/0908.4052}{{\tt arXiv:0908.4052}}.
	
	\bibitem{Bonelli:2025owb}
	G.~Bonelli, A.~Shchechkin, and A.~Tanzini, {\it {Refined Painlev\'e/gauge
			theory correspondence and quantum tau functions}},
	\href{http://arxiv.org/abs/2502.01499}{{\tt arXiv:2502.01499}}.
	
	
	
	\bibitem{Matone:1995rx}
	M.~Matone, {\it {Instantons and recursion relations in N=2 SUSY gauge theory}},
	{\em Phys. Lett.} {\bf B357} (1995) 342--348,
	[\href{http://arxiv.org/abs/hep-th/9506102}{{\tt hep-th/9506102}}].
	
	\bibitem{Fioravanti:2019vxi}
	D.~Fioravanti and D.~Gregori, {\it {Integrability and cycles of deformed ${\cal
				N}=2$ gauge theory}},  {\em Phys. Lett. B} {\bf 804} (2020) 135376,
	[\href{http://arxiv.org/abs/1908.08030}{{\tt arXiv:1908.08030}}].
	
\end{thebibliography}
\providecommand{\href}[2]{#2}\begingroup\raggedright\endgroup

\end{document}